\documentclass{article}
\usepackage[scaled]{helvet}

\usepackage[english]{babel}

\usepackage[letterpaper,top=2cm,bottom=2cm,left=3cm,right=3cm,marginparwidth=1.75cm]{geometry}

\usepackage{amsmath}
\usepackage{graphicx}
\usepackage[dvipsnames]{xcolor}
\usepackage[colorlinks=true, allcolors=teal]{hyperref}
\usepackage{xspace}
\usepackage{csquotes}
\usepackage{setspace}
\usepackage[round]{natbib}
\newcommand{\kmer}{$k$-mer\xspace}
\newcommand{\kmers}{$k$-mers\xspace}
\title{Advances in colored \emph{k}-mer sets: essentials for the curious}

\begin{document}
\author{Camille Marchet $^1$ \\$^1$ UMR9189 CRIStAL, Univ Lille, CNRS, Centrale, F-59000 Lille, France\\camille.marchet@univ-lille.fr}
\date{}
\maketitle

\begin{abstract}
     This paper provides a comprehensive review of recent advancements in \kmer-based data structures representing collections of several samples (sometimes callded colored de Bruijn graphs) and their applications in large-scale sequence indexing and pangenomics. The review explores the evolution of \kmer set representations, highlighting the trade-offs between exact and inexact methods, as well as the integration of compression strategies and modular implementations. I discuss the impact of these structures on practical applications and describe recent utilization of these methods for analysis. By surveying the state-of-the-art techniques and identifying emerging trends, this work aims to guide researchers in selecting and developing methods for large scale and reference-free genomic data. For a broader overview of \kmer set representations and foundational data structures, see the accompanying article on practical \kmer sets.
 \end{abstract}
%
%%%%%%%%%%%%%%%%%%%%%%%%%%%%%%%%%%%%%%%%%%%%%%%%%%%%%%%%%%%%%%%%%%%%%%%%%%

%you may choose to use a table of content
%\tableofcontents
\onehalfspacing
\section{Introduction}
\subsection{Context}
\begin{figure}[ht]
    \centering
    \includegraphics[width=0.8\textwidth]{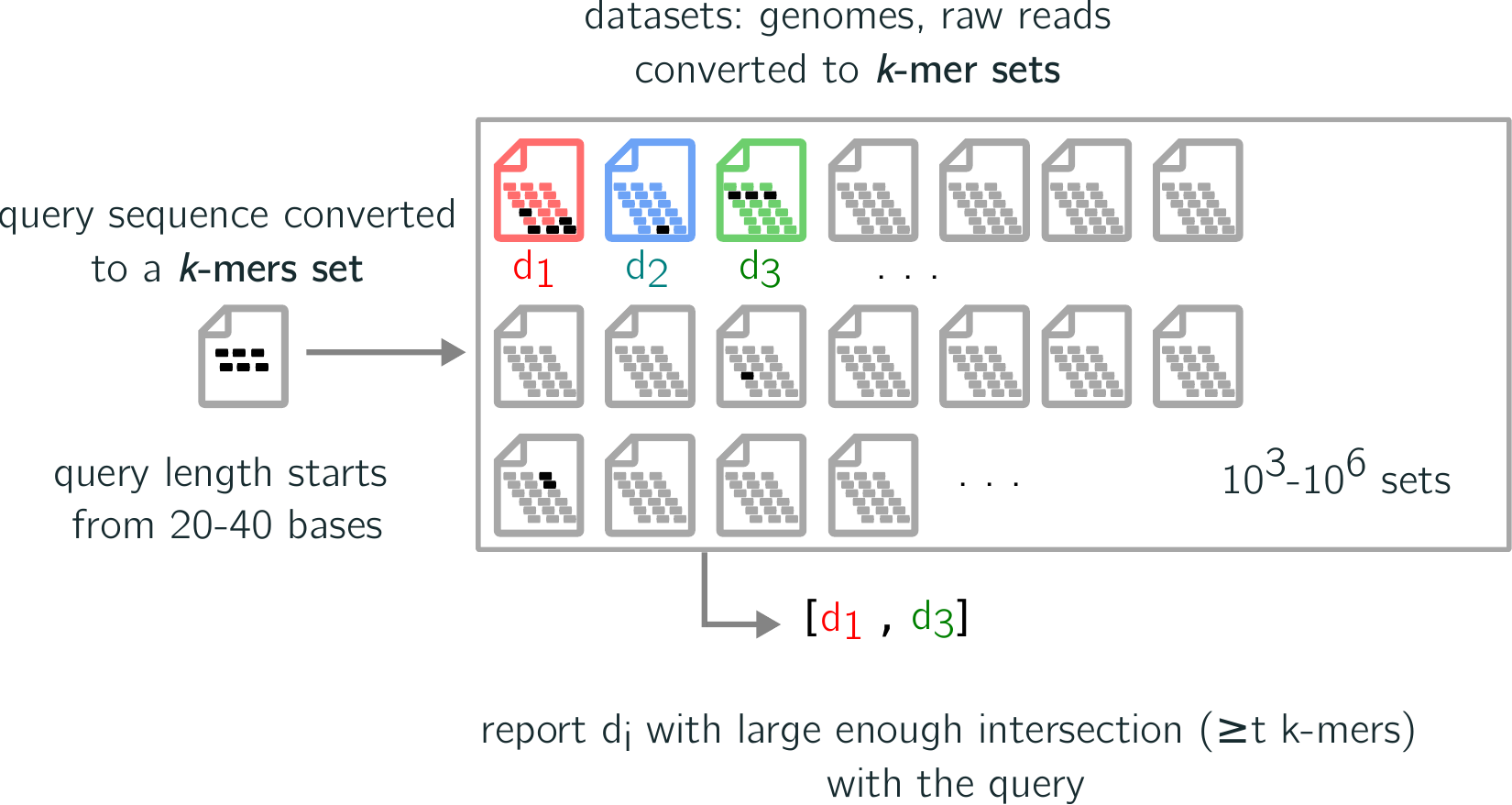}
    \caption{Illustration of the indexed data and basic query using a colored \emph{k}-mer set structure.}
    \label{fig:intro_question}
\end{figure}

\begin{figure}[ht]
    \centering
    \includegraphics[width=\textwidth]{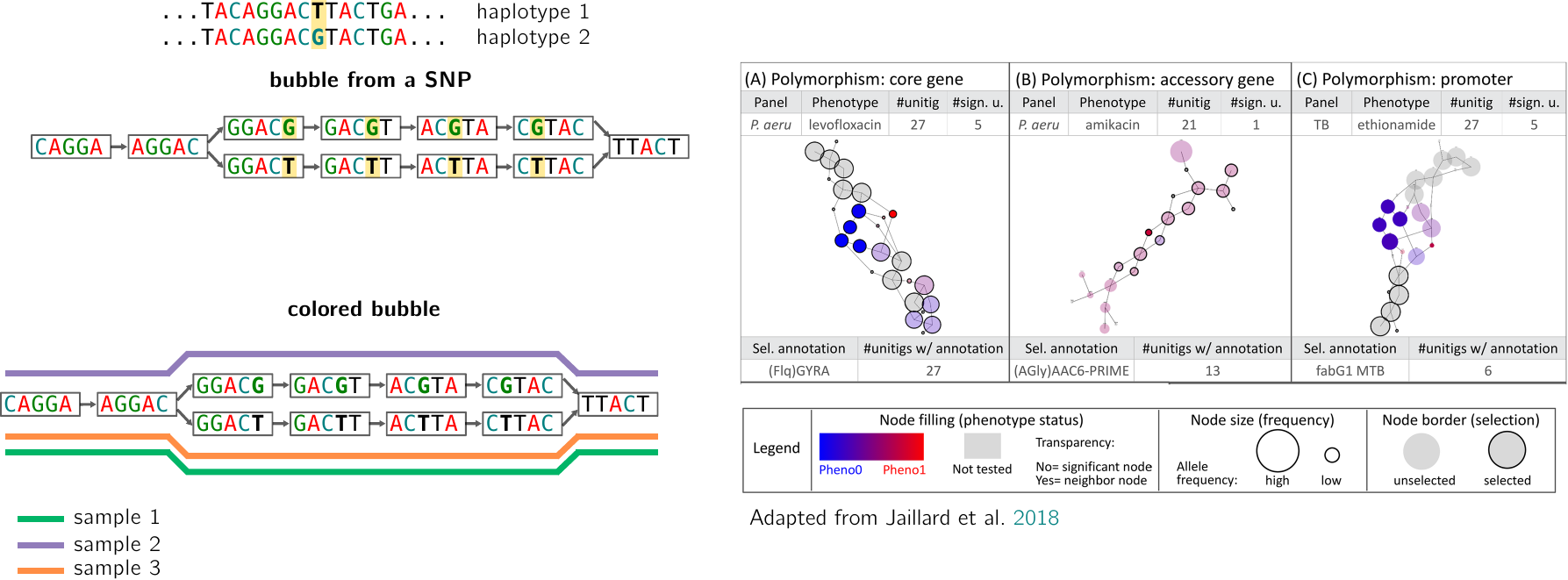}
    \caption{A bubble is a subgraph that represents alternative paths between a source and a sink node. The subgraph starts at a source node that is common to all paths, and ends at another common sink node, with multiple divergent paths in between that eventually converge at the sink. Bubbles typically represent sequence variations, such as SNPs (single nucleotide polymorphisms), insertions, deletions or alternative splicing events, or more complex events, in the graph.\\
    Left: a bubble induced by a SNP in a de Bruijn graph. Left bottom: a bubble whose paths exist differentially in 3 samples. Unitigs for that graph would be "CAGGAC", "GGACGTAC", "GGACTTAC" and "TTACT".  The graph is colored with the samples. Right: adapted from \cite{jaillard2018fast}, an example case of using the de Bruijn graph's bubbles to observe phenotypes \textit{P. aeruginosa} population. Here the graph nodes are simplified (unitigs). Colours are a spectrum between blue for susceptible nodes and red for levofloxacin resistance.}
    \label{fig:dbg_bubble}
\end{figure}

In the realm of pangenomics, the use of graph structures is predominantly associated with variation graphs such as PGGB~[\cite{garrison2023building}] and minigraph cactus~[\cite{hickey2024pangenome}], or with graphs of homologous genes in microbial pangenomics~[\cite{colquhoun2021pandora}]. Recent advancements have introduced structures based on full-text index improvements, such as the r-index (e.g., SPUMONI2~[\cite{ahmed2023spumoni}]) and the Move index (e.g., Movi ~[\cite{zakeri2023movi}]). These structures scale linearly in space with the number of distinct sequences in the dataset and have made significant progress in efficiency and query performance. Another line of work in the literature have introduced various proposals based on \kmers (genomic words of size $k$) and a graph of sequences initially introduced to assemble short reads, the de Bruijn graph. Contrary to pangenome structures mentioned before, many of these structures can also deal with unassembled data with little or no pre-processing. They allow to study genomic variations across multiple datasets simultaneously by assigning each dataset a "color" within the graph. This innovation enabled the identification of shared and unique sequences (see Figure \ref{fig:intro_question}), as well as the detection of variations like SNPs (see Figure \ref{fig:dbg_bubble}),directly from sequencing data.\\

At the core of these structures are decisions regarding internal \kmer representation, that are tied to performance trade-offs, such as which operations can be offered and at what time/memory cost. Common operations include lookup, navigation within the de Bruijn graph, ranking, set operations (union, intersection, difference), and the addition/deletion of elements. The \kmer based strategy has recently been pointed out as an interesting option for scalability in pangenomics [\cite{andreace2023comparing}]. It is also an asset for studying species with poorly known genomes or non-reference variants [\cite{krannich2022population}], and is explored for transcriptomes [\cite{morillon2019bridging}]. Recent colored structures are also used in updated versions of well-established tools, such as Kallisto [\cite{sullivan2023kallisto}] for quantification of gene expression with RNA-seq. 
This paper addresses \kmer-based structures, particularly de Bruijn graphs, and reviews advances in large scale indexes for collections of sequence, as a follow-up of~\cite{marchet2021data}. I will overview novel techniques that emerged since 2020, novel lines of works and trends, as well as applications that developed in the recent years.

\subsection{Preliminaries}
\subsubsection{\emph{K}-mer sets and de Bruijn graphs}
In genomics, a \kmer is a substring of length k derived from a biological sequence, such as DNA or RNA. For example, the sequence "AGCT" has 3-mers "AGC," "GCT," and "CTA." $K$-mers are often represented as binary strings or integers, allowing for efficient storage and querying. The choice of k significantly impacts the trade-offs in computational performance, with smaller values of k providing more generality, and larger values offering higher specificity.

A de Bruijn graph is a data structure that represents overlaps between sequences in a compact form. In the context of genomics, it is constructed using \kmers as nodes, with directed edges indicating overlaps of k-1 bases between \kmers. For example, if the \kmer "AGC" is followed by "GCT" in the sequence, there is an edge from the node representing "AGC" to the node representing "GCT." Therefore, a \kmer sets always implicitely represents a de Bruijn graph. De Bruijn graphs also highlight genomic variations in patterns (see Figure \ref{fig:dbg_bubble}).

From a de Bruijn graph, several objects can be constructed.\\ \textbf{Unitigs} are simple paths in a de Bruijn graph, where each node has an in-degree and out-degree of one (find an example in Figure \ref{fig:dbg_bubble}). Unitigs are the building blocks of more complex paths and are often used as the basic elements in genome assembly.\\
\textbf{Minimizers:} Minimizers are a technique used to choose representative sequences of a \kmer set according to a predetermined rule, such as lexicographical order or hash. This sampling strategy helps in optimizing storage through partitioning and speeding up computations (Figure \ref{fig:basic} (g,h,i)).\\
\textbf{Super-\kmers:} These are sequences that concatenate consecutive \kmers sharing the same minimizer. Super-\kmers help to further compress \kmer sets and reduce redundancy, which is particularly useful in large-scale genomics data processing (Figure \ref{fig:basic} (h)).\\
More generally, \textbf{spectrum preserving string sets} (SPSS) are sequences longer than \kmers that aggregate \kmers similarly to unitigs or super-\kmers. They are built from a de Bruijn graph and allow to co-encode \kmers, reducing the cost of representing a set.

\subsubsection{Basic data structures}
\begin{figure}[ht]
    \centering
    \includegraphics[width=\textwidth]{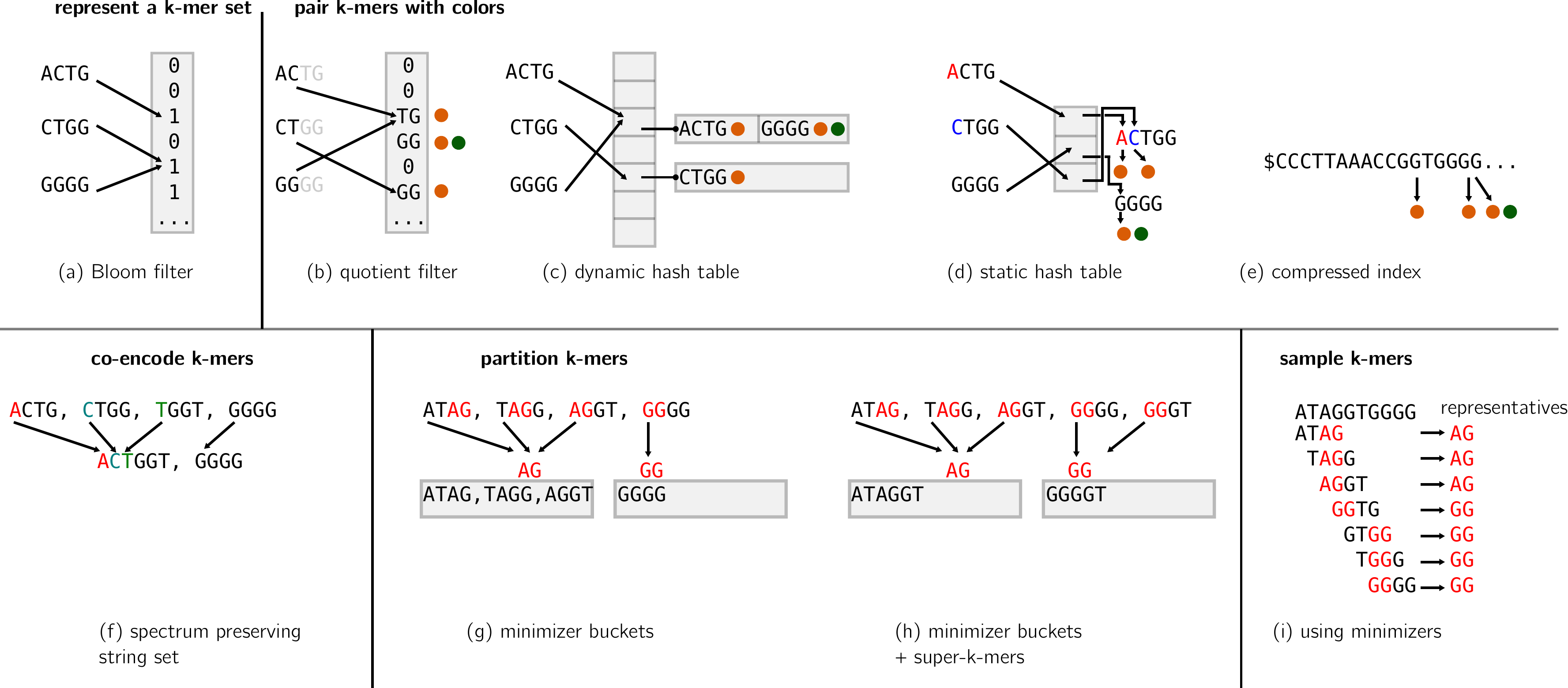}
    \caption{Basic structures and strategies to work with \kmer sets. Bloom filters (a) allow membership queries of \kmers and structures (b,c,d,e) can also associate \kmers with information. (f) Spectrum preserving string sets (SPSS) are sets of strings from which an input set of \kmer can be spelled. Here, \kmers are assembled in unitigs, which are an example of SPSS. In (g,h,i) minimizers are represented in red. In these examples, they are the smallest lexicographic \emph{m}-mer (\emph{m=2}) found in \kmers. \emph{K}-mers sharing a minimizer can be grouped in similar buckets for partition (g), and can be further compacted in SPSS in these buckets (h). Spectrum preserving string sets created from consecutive \kmers sharing a minimizer as in (h) are called \emph{super-\kmers}. (i) Minimizer can also be a way to sample \kmer sets. }
    \label{fig:basic}
\end{figure}

$K$-mer sets can be represented and indexed in different types of data-structures.
They sometimes differ by fundamental properties, such as being an exact representation, or an inexact one (often with false positives when querying the structure). Inexact representations that matter to us include:
\begin{itemize}
    \item[$\diamond$] Bloom filters. A Bloom filter is a space-efficient probabilistic data structure used to quickly test whether an element is part of a set, with the trade-off that it may produce false positives but never false negatives. A Bloom filter works by hashing the input element multiple times and setting the corresponding bits in a bit array to 1; to check for membership, it verifies that all these bits are set to 1, which indicates possible presence, though with a chance of false positives (Figure \ref{fig:basic} (a)).
    \item[$\diamond$] Quotient filters. A structure close to Bloom filters but that manages hash collisions (Figure \ref{fig:basic} (b)). The counting quotient filter variant can keep track of the count of each item.
\end{itemize}
They all represent \kmers as integer signatures or bits in a table. To transform \kmers into integers, they rely on hash functions, that also help them to populate the table harmoniously. 

Exact methods include methods that see \kmers as hashes and methods that leverage lexicographic properties of the \kmer set, and sometimes a mix of the two:
\begin{itemize}
    \item[$\diamond$] Hash tables. They are map structure associating and storing (key/value) pairs using hash functions). Space efficient hash table rely on different optimization specific to \kmer sets to gain space and speed-up queries (Figure \ref{fig:basic} (c) and (d)).
    \item[$\diamond$] Compressed lexicographic indexes. They allow to query a compressed set of \kmers based on the Burrows Wheeler transform (BWT)  (Figure \ref{fig:basic} (e)).
    \item[$\diamond$] Lexicographic tree structures. They leverage the redundancy in the \kmers to organize them by common suffixes or prefixes in a hierarchical manner. 
    \item[$\diamond$] Filters if \kmer fingerprints are lossless (Figure \ref{fig:basic} (b)).
\end{itemize}

I described all these structures and their use for \kmer sets more comprehensively in the companion paper on \kmer set structures [\cite{Marchet2024}], they were previously reviewed in \cite{chikhi2021data}.

\subsubsection{Colors}
Colored \kmer sets aggregate different datasets, allowing comprehensive querying across multiple read sets or genomes. Membership queries return datasets intersecting significantly with query \kmers, and other types of operations are now permitted with some of these methods, such as pseudo-mapping in~[\cite{fan2024fulgor}].

Color can be a confusing terminology with discrepancies across papers. A majority of papers refer to a color to denote individual paths with \kmers pertaining to a dataset. Then, efficient storage uses color sets, representing \kmers with the same dataset set. These color sets are sometimes themselves called colors, hence the confusion. A color could in fact be any information that can label a path. Colors have been called bins [\cite{mehringer2023hierarchical}], and colored structures have also been called tinted de Bruijn graph [\cite{vandamme2024tinted}] or counting de Bruijn graph [\cite{karasikov2022lossless}].
In this survey, we use \emph{color} to refer to individual integers associated with \kmers, and \emph{color set} to refer to the set of integers associated with a \kmer (e.g. the set of datasets it is contained in).

\begin{figure}[ht]
    \centering
    \includegraphics[width=\textwidth]{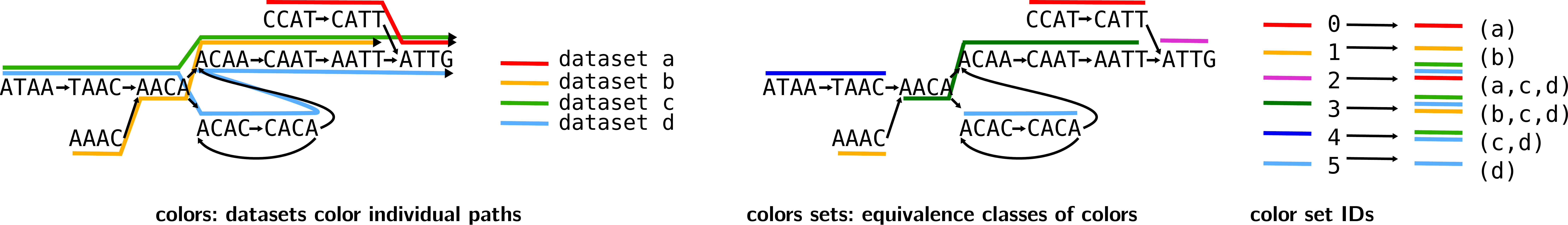}
    \caption{Illustrations of colors in de Bruijn graphs of biological sequences.}
    \label{fig:def_colors}
\end{figure}
\begin{figure}[ht]
    \centering
    \includegraphics[width=1.1\textwidth]{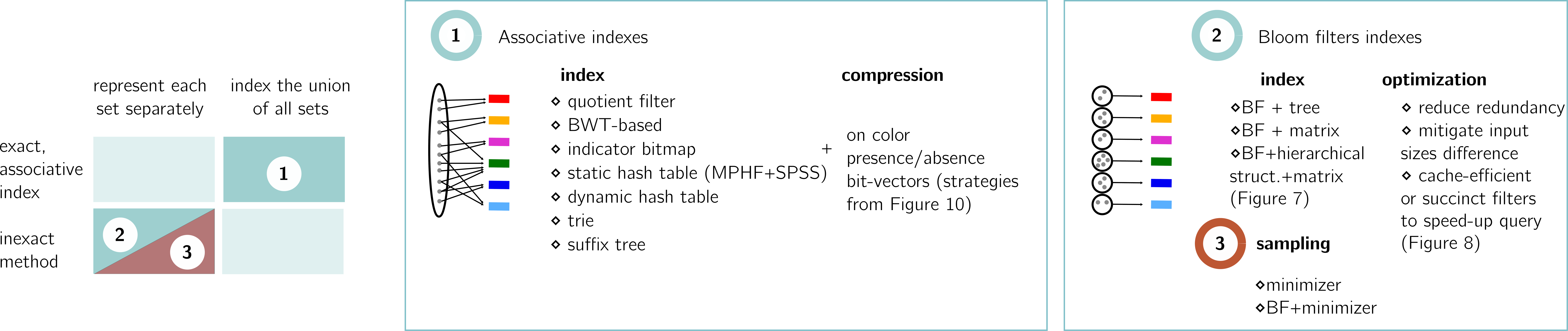}
    \caption{Categorization of the state-of-the-art for colored \kmer sets}
    \label{fig:sota_colors}
\end{figure}

\begin{figure}[ht!]
    \centering
    \includegraphics[width=\textwidth]{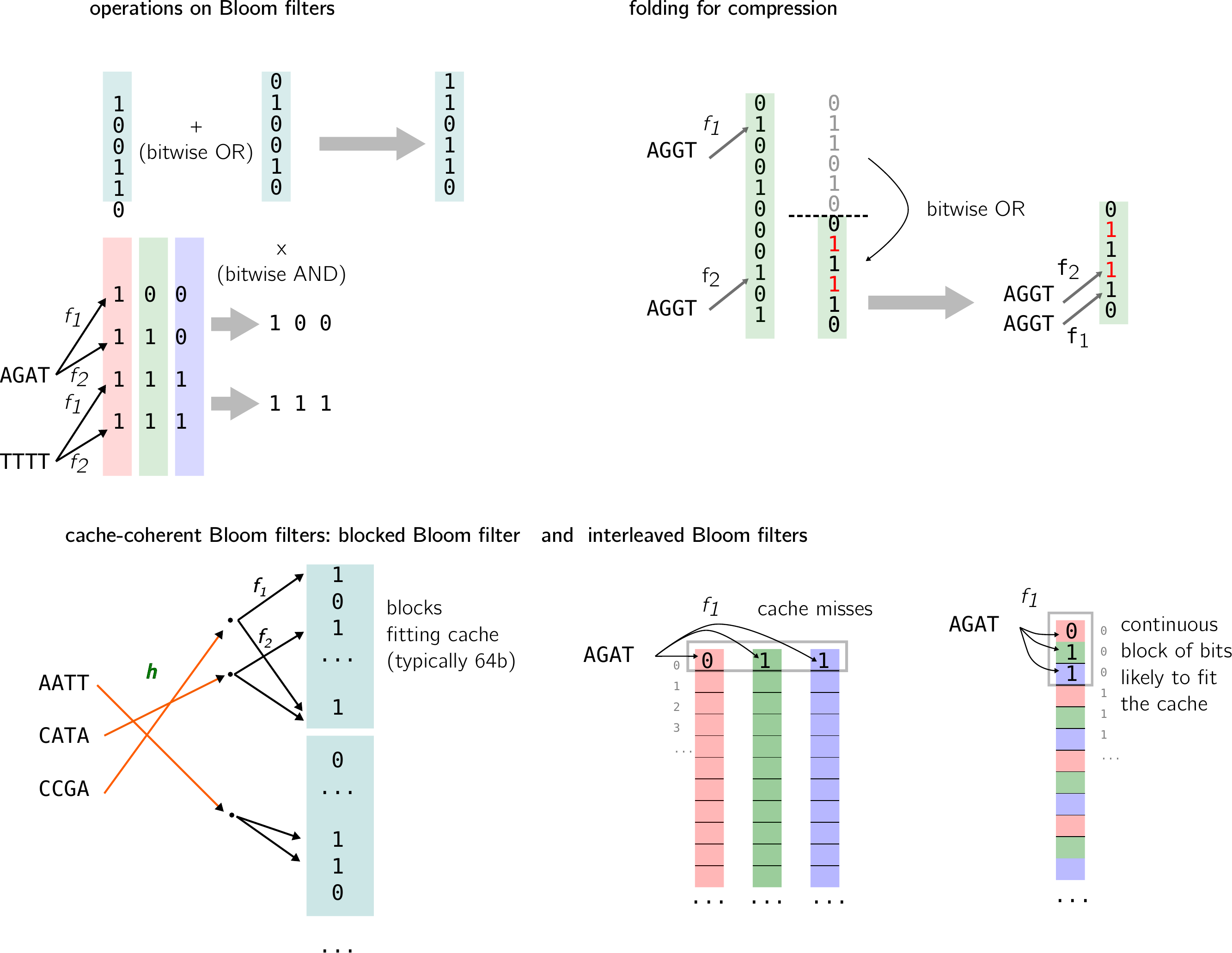}
    \caption{Operations and optimizations of Bloom filters. The bitwise OR operation is used for merging filters, for instance in tree based inner nodes or root construction, or when folding filters. The bitwise AND operation is used to verify if all assigned bits of a queried \kmer attest for its presence.\\
    Bloom filters can be folded for compression, typically by merging one half with the other in a smaller filter.\\
    Modifications of Bloom filters aimed at improving query speed by making the filters cache efficient. Blocked Bloom filters are described in \cite{Marchet2024}, and organise the \kmer's bit so that they can all be checked in a cache-fitting block.\\
    Interleaved Bloom filters act on several Bloom filters, and make the simultaneous search of a \kmer in all these filters quicker. They are useful when the filters have primarily been constructed with equal settings, so the search of a particular \kmer hits same addresses in different filters.
    Regular search implies loading distant parts of different filters. Instead, bits associated to a \kmer are interleaved in a contiguous block of a single filter, making the query more likely to fit the cache.}
    \label{fig:operation_bf}
\end{figure}
Colored datasets inherit from the duality of exact/inexact structures for \kmers. A state-of-the-art categorization is therefore exact versus inexact methods. A vast majority of exact methods use a union set of the \kmers from all datasets (reviewed in section \ref{section:exact}). Exact indexes link \kmers to presence/absence vectors across datasets, compressing color information for efficient access. At the opposite, inexact methods index separately each dataset.

Inexact methods represent each set separately using Bloom filters, associating filters with colors via trees or matrices (reviewed in section \ref{section:inexact}). 

In the following, I will focus on structures in which the query is quite precise and possible for a single \kmer. I won't review structures meant for genome scale queries on large collections based on MinHash, that have been described in \cite{rowe2019levee}.

\section{Inexact methods}\label{section:inexact}

\subsection{Index all \emph{k}-mers with false positives or false negatives}
Most inexact methods reviewed here store \kmers use Bloom filters for each set.
On top of Bloom filters, inexact methods avoid independent queries for each set via tree structures or matrices. 
In both matrix and tree strategies, Bloom filters have the same size and hash functions for all datasets, and must be parameterized adequately. 

Figure \ref{fig:operation_bf} describes union of Bloom filters, that is used to build tree-like indexes. Union is realized through a bitwise or that compares two bits at the same position in two bit arrays and returns 1 if at least one of the bits is 1, otherwise 0.\\
Figure \ref{fig:operation_bf} also shows bitwise and operation that compares two bits at the same position in two bit arrays and returns 1 only if both bits are 1, otherwise 0. This operation is used to check if a \kmer is present in a a filter.

\begin{figure}[ht]
    \centering
    \includegraphics[width=\textwidth]{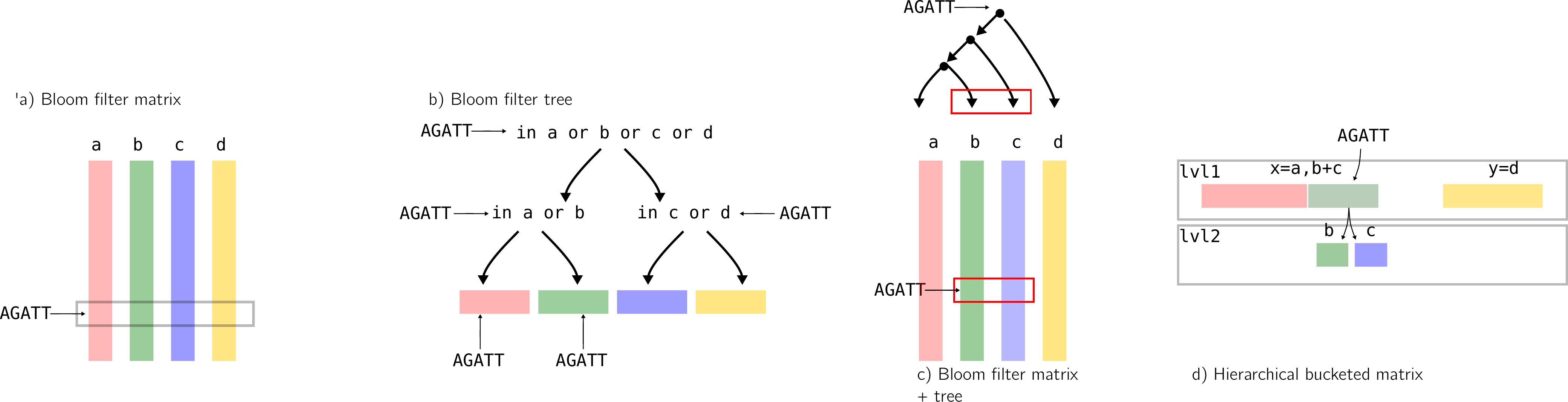}
    \caption{Main strategies for inexact methods. (a) and (b) were reviewed in \cite{marchet2021data}, and (c) and (d) represent improvements. In (a), Bloom filters per dataset are stacked as columns of a matrix, and the queries are performed on rows. In (b), filters are aggregated in a tree structure, and starting from the root, a \kmer's presence is searched in subtrees of nodes where it is present (for instance, AGATT is not present in node (c or d) and therefore not looked up in blue or yellow filters). In (c), a comb structure helps querying more precise slices of a matrix of Bloom filters, than queries in (a). In (d), filters are stacked to fill bins, and a hierarchical structures descends to retrieve precisely the initial datasets of a \kmer.}
    \label{fig:probabilistic}
\end{figure}

\subsubsection{Tree strategies} They were among the first proposed as a solution for the colored \kmer set indexation problem~[\cite{solomon2016sbt}]. In the tree strategy, trees union Bloom filters (Figure \ref{fig:operation_bf} shows union) at each level. The root of the tree represents all \kmers pertaining to any indexed datasets, and each leaf is a dataset (Figure \ref{fig:probabilistic} (b)). Different improvements have worked on better fill-up for the structure, and on reducing the redundancy brought by unions at each level [\cite{solomon2018ssbt,sun2017allsome}], the most recent being HowDeSBT [\cite{harris2018howdesbt}]. Instead of knowing whether a \kmer is somewhere in a subtree when querying a node, HowDeSBT informs as early as possible whether a \kmer is present everywhere or absent everywhere in a subtree.
In these structures, a query proceeds by starting at the root and descends in subtrees where the \kmer is marked as present in the filters, until leaves are reached. 

\subsubsection{Matrices strategies} They stack filters in inverted matrices, querying presence or absence in all datasets through a single row using \kmer hashes. The final presence/absence vector used rapid bit summation to report the hits in each sample (Figure \ref{fig:probabilistic} (a)). The technique was first proposed in Dream-Yara [\cite{dadi2018dream}] and BIGSI [\cite{bradley2019ultrafast}] was the first method of this kind and with variations in several other works [\cite{Seiler2021,10.1093/bioinformatics/btad101,10.1093/bioinformatics/btac845}].
BIGSI reimplementation in COBS [\cite{Bingmann2019}] allow to resize a fraction of the filters to mitigate datasets varying by order of magnitudes their \kmer cardinalities, and to compress chunks of Bloom filters.

\subsubsection{Other filters}
A distinct structure relies on a hashing strategy similar to minimal perfect hashing (see subsection \ref{section:hash_tables}) and to other static filters (refer to \cite{Marchet2024}), called Othello hashing. The tool, SeqOthello [\cite{yu2018seqothello}], associates \kmers to colors with possible false positives using an MPHF and small fingerprints. Contrary to other inexact methods, and similarly to the exact method's approach, it associates the union of \kmers from all samples to their color vectors.

\subsubsection{Sampling structures}
Recent indexes also made the choice to sample the indexed \kmers to scale to the whole sequenced nucleotides. It allows for the selection of \kmers representative based on a predefined rule, further optimizing the indexing process by concentrating on the most informative sequences. These sampling strategies provide scalable solutions for handling the increasing volume of sequencing data in genomics and are based on minimizers (see \ref{fig:basic} (i)).\\

Needle [\cite{darvish2022needle}] is based on interleaved Bloom filters (see Figure \ref{fig:operation_bf}). Instead of indexing \kmers, it samples the data by harvesting minimizers (Figure \ref{fig:basic} (i)), and associates them to counts using a variant of Bloom filters (possibly, counts can be a bit over-estimated).\\
Pebblescout  [\cite{shiryev2024indexing}] is based on a structure that highly optimizes tries. It also partitions minimizers based on their integer representation and an encoding technique used in cryptography that aims at shuffling \kmers in partitions deterministically.

\subsubsection{Errors in queries}
In inexact structures, queries are usually several \kmers long. Therefore, the important matter is the false positive rate of the entire set of queried \kmers. When querying multiple \kmers, each \kmer is checked individually. Even if a single \kmer results in a false positive, the other \kmers in the query can still provide correct information. This means that the likelihood of the entire query producing a false positive is significantly lower than the false positive rate for a single \kmer.\\ Despite not indexing \kmers, Pebblescout handles queries starting from 42 nucleotides, because for each 42-mer at least one minimizer 25-mer is saved in the index.

\subsection{Optimizing the inexact structures}

In structures based on stacked Bloom filters (both in trees or matrices), query consistency and efficiency across the different columns of the matrix or levels of the tree is ensured by always using the same hash functions and Bloom filter sizes for each dataset.\\
An unbalanced size distribution of indexed samples poses significant challenges. This occurs because the Bloom filter size and hash functions are calibrated for a false positive rate based on the largest dataset. If the structure contains a mix of small and large filters, the large filters can impose a waste of space on the smaller ones. In the following, I show how novel improvements tackled this issue.\\
Descending in a tree or jumping in a large matrix can quickly be time inefficient for a large enough input datasets. I also show how structures work on query speed.

\subsubsection{Query speed}
In theory, tree strategies have a better query time complexity (sublinear in the number of input datasets) in the best case. Up to now, matrices have proven practically capable to optimize space and cache efficiency, and preferred to trees in recent works. PAC [\cite{10.1093/bioinformatics/btad225}] proposed a hybrid approach that re-organize a tree to make the search cache-efficient and to go directly in relevant slices in an inverted matrix (Figure \ref{fig:probabilistic_comp}(e)). In the most favorable case (rare \kmers), this structure has a constant response time.\\
Several works such as MetaProFi [\cite{10.1093/bioinformatics/btad101}] rely on dividing Bloom filters into several parts, or interleaved to optimize query efficiency. By chunking Bloom filters, one can load separate parts into memory to reduce the overhead and use parallel processing~[\cite{10.1093/bioinformatics/btad101}]. For instance, Blocked Bloom filters or interleaved Bloom filters are designed to improve cache efficiency by organizing the Bloom filter into blocks that fit well into the CPU cache lines (Figure \ref{fig:probabilistic_comp}(d)). This organization minimizes the number of cache misses during querying, as all relevant data for a query is likely to be found within the same cache block. Another solution, not pictured in Figure \ref{fig:probabilistic_comp}, completes PAC's approach. The structure is partitioned according to minimizers, each minimizer gets a small Bloom filter for its \kmers. That way, consecutive \kmers are more likely to be queried in a cache-coherent way as a single, small filter is loaded.\\
COBS [\cite{Bingmann2019}] uses SIMD (a parallel computing technique that allows a single instruction to simultaneously perform the same operation on multiple data elements) to speed up the querying process. During a query, COBS needs to perform multiple bitwise operations to check whether a \kmer is present in the indexed data. By using SIMD, COBS can check multiple bit positions at once across different slices of the signature index, rather than performing these checks sequentially.

\subsubsection{Mitigate variable input sizes}

\begin{figure}[ht]
    \centering
    \includegraphics[width=\textwidth]{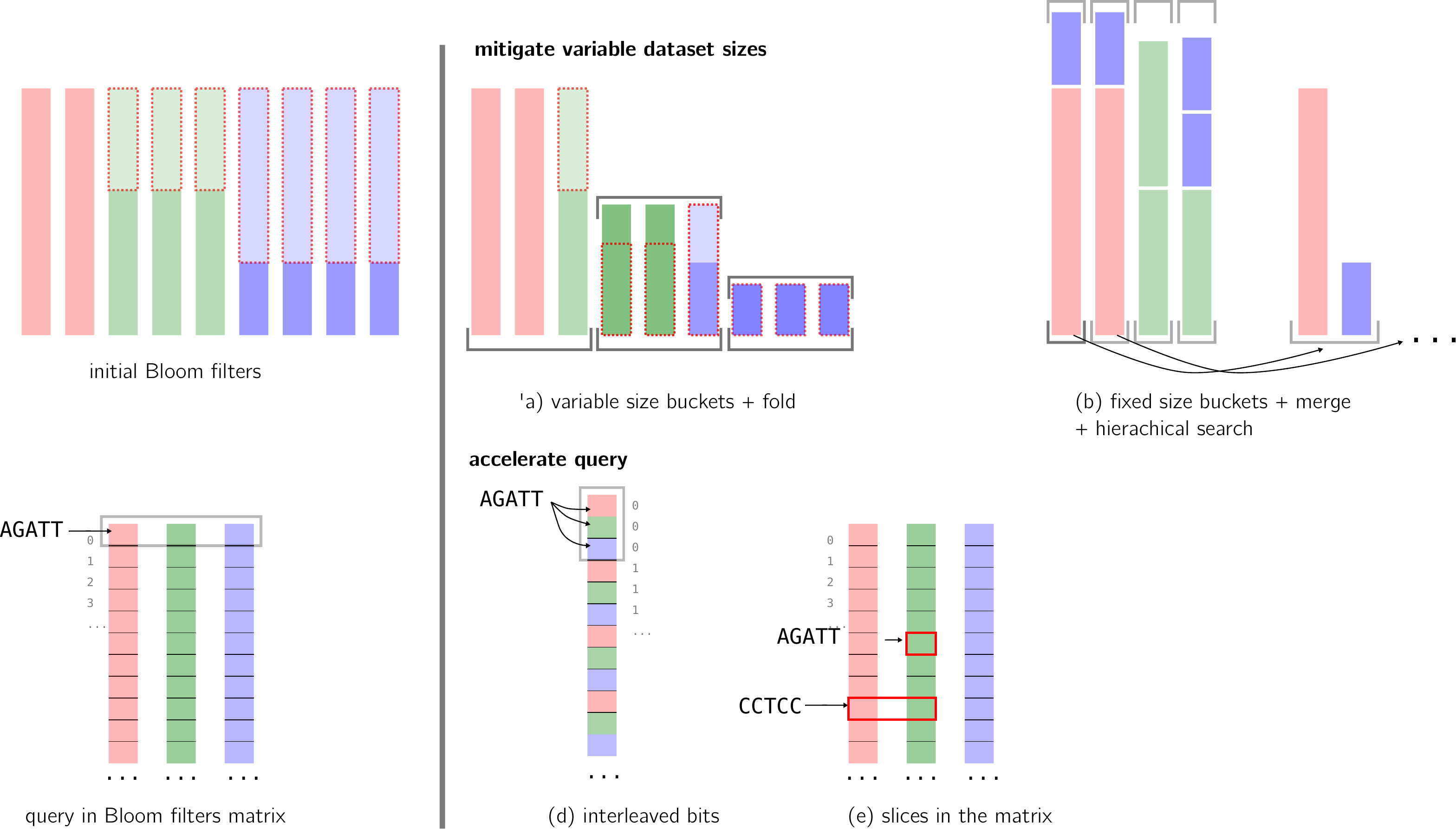}
    \caption{Compression and improvement of query speed in inexact methods. Initial Bloom filters for a colored set are represented on the left. They are dimensioned according to the filters containing the most \kmers (red ones). In the other ones, a waste of space is represented by a dashed region. This issue is mitigated through strategies (a) and (b). In (a), Bloom filters are folded (see Figure \ref{fig:operation_bf}) once or several time to create buckets that fit better the efficient size each filter should require. In (b) initial Bloom filters are stacked to attain a fixed size merged size, and a hierarchical structure allows to retrieve single filters when \kmers are queried. To accelerate queries in a matrix of Bloom filters, works have relied on interleaving bits (d) of the collection, or in querying restrained slices of the matrix (e).}
    \label{fig:probabilistic_comp}
\end{figure}

When working with Bloom filters in blocks or slices, some matrices approaches apply run-length encoding (compress data by representing sequences of repeated characters as a single character followed by the number of its occurrences).
COBS proposed to use a discrete size categorization with several bins containing Bloom filters of same size. Originally, all Bloom filters are created equal, and if needed their size can be reduced using folding (representing several filters at once, or a single filter in half its size, using a bitwise OR operation). This operation increases the false positive rate of the filter, but filters can all still be queried with the initial hash functions despite the resizing (Figure \ref{fig:probabilistic_comp}(a)).\\
Raptor [\cite{Seiler2021}] implements a hierarchical interleaved Bloom filters (HIBF [\cite{mehringer2023hierarchical}]) strategy. It relies on fixed size buckets, this time by merging Bloom filters of different sizes to obtain a filter of the required size. It requires a hierarchical structure to know in detail the original datasets of a \kmer contained in a merged filter (Figure \ref{fig:probabilistic_comp}(b)).
A query is first checked against the higher levels of the hierarchy. If a potential match is found, the query is then passed down to the next, more detailed level for further checking. Filters are also organized in a cache efficient way.

\subsubsection{Scalability}
One recent line of development memory mapped inverted matrix~[\cite{Lemane2024}], a data structure that represents the layout of memory addresses directly on the disk. It avoids important RAM costs and still provides fast queries using partitioned \kmer bins based on minimizers. This results in an index whose limit is the available disk, that can therefore be very scalable. COBS also proposes a memory-mapped version.

\section{Exact methods}\label{section:exact}

\subsection{Overview of exact methods}
Exact methods rely on a \kmer set structure to represent the union of all distinct \kmers of the collection. Each \kmer is linked to a row in a presence/absence matrix with datasets in columns, on which a compression strategy is applied. Exact methods use an associative index, that pairs each \kmer with a specific value such as its color.

\subsubsection{Methods using hash tables}\label{section:hash_tables} The first generation of structures used regular hash tables [\cite{iqbal2012novo}] (see Figure \ref{fig:basic} (c)), and efficient dynamic hash tables remain used in several works (Bifrost [\cite{holley2019bifrost}], Metagraph [\cite{Karasikov2020.10.01.322164}]), notably for their dynamic properties.

Nowadays, other methods opted for efficient \kmer hashing using whose gain in space is brought by minimal perfect hash functions, co-encoded \kmers (Pufferfish [\cite{almodaresi2018pufferfish}]),
and partitioning (Fulgor and Mac-dBG [\cite{Pibiri2024,fan2024fulgor}], GGCAT [\cite{cracco2023extremely}], Reindeer [\cite{marchet2020reindeer}].
They rely on two key components: \emph{minimal perfect hashing} that allows to allocate a minimal amount of space regarding the input \kmer set, and a technique for co-encoding key values that relies on assembled paths in de Bruijn graphs (\emph{spectrum preserving string sets}).

Main improvements in this category of methods have been the combination of new spectrum preserving string sets for better handling key \kmers, of cache-efficient minimal perfect hash functions and of advanced color compression strategies (see subsection \ref{section:compression}). Fulgor [\cite{fan2024fulgor}] is a good example. Other advances include storing integers instead of presence/absence, for the representation of counts or positions [\cite{marchet2020reindeer,karasikov2022lossless}].

\subsubsection{Methods using filters} They often rely on the associativity property of quotient filters for recording counts or more generally integers (Mantis [\cite{Pandey2018}] dynMantis [\cite{almodaresi2022incrementally}]). They have a larger space footprint than their counterpart inexact version where \kmer signatures are lossy.\\

These structures mainly improved toward the capacity to insert new datasets [\cite{almodaresi2022incrementally}], and motivated works on color-compression.

\subsubsection{Methods using compressed lexicographic indexes} They integrate a lexicographic compression of their \kmer set combined with an index (Rainbowfish [\cite{almodaresi2017rainbowfish}], VARI and VARI-merge [\cite{muggli2017vari,muggli2019building}], Metagraph [\cite{Karasikov2020.10.01.322164}], Themisto [\cite{alanko2023themisto}]).
More precisely, they rely on variants of the original BWT that has specialized for \kmer sets and refined through time to represent succinctly the de Bruijn graph. Their strength is space efficiency.

Recently, developments have made them competitive for single \kmer queries speed [\cite{alanko2023longest}], and combined a new space efficient representation of de Bruijn graphs to color-compression strategies [\cite{alanko2023themisto}].

\subsubsection{Other method} Methods using tree structures can be found mixed with other data structures [\cite{marcus2014splitmem,agret2020redoak}], such as prefix trees storing Bloom filters at each node instead of \kmers themselves and handling false positives \cite{holley2016bftrie}.
Methods using data-bases (Pantools [\cite{10.1093/bioinformatics/btw455}]) opt for graph-specialized database managers, such as Neo4j, to store a colored de Bruijn graph associated to meta-data.

\subsection{Modularity in exact methods}
\begin{figure}[ht]
    \centering
    \includegraphics[width=\textwidth]{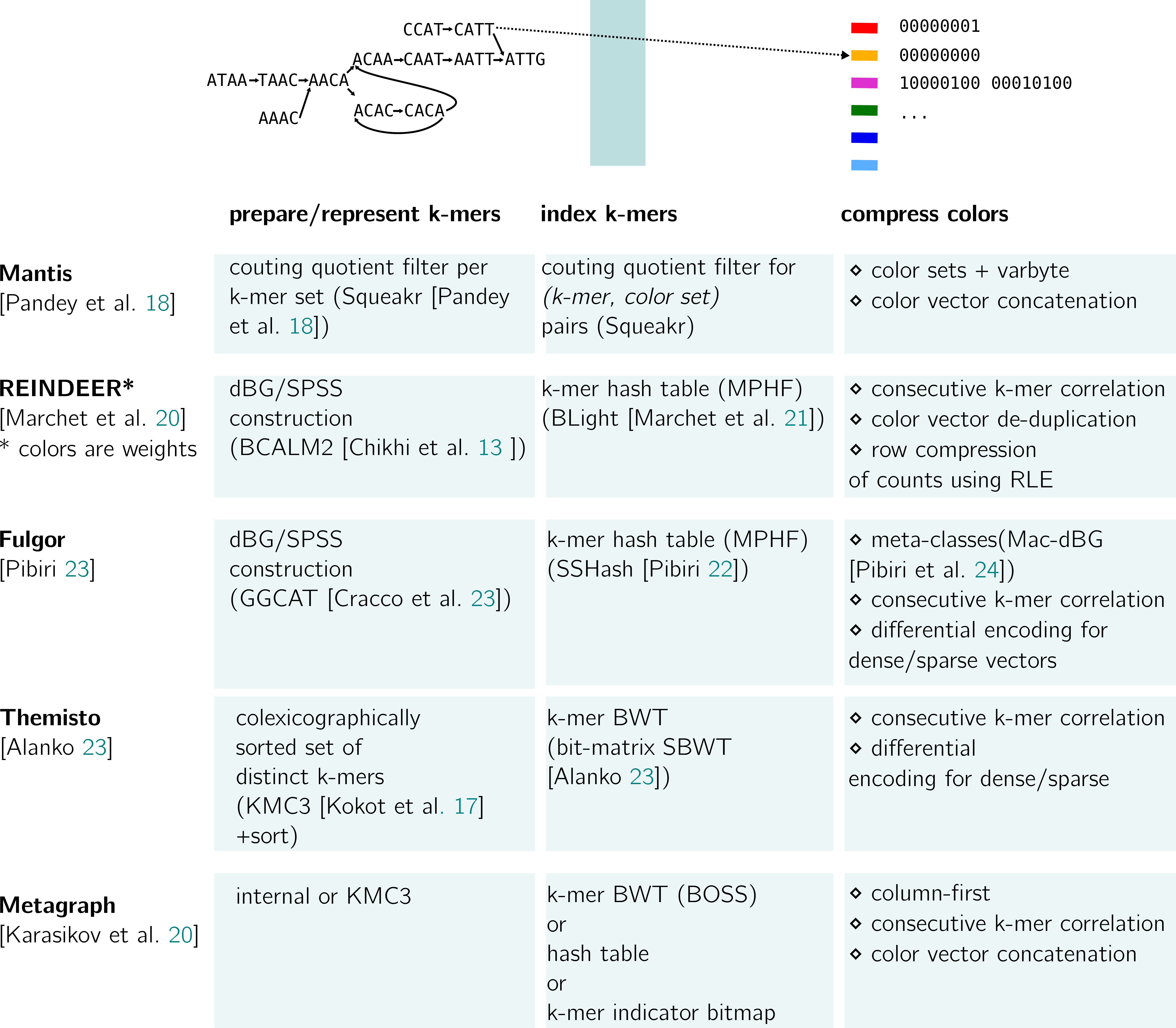}
    \caption{Examples of modularity in exact methods for indexing \kmers. See Figure \ref{fig:compression_strat} for a description of compression strategies. Reindeer compresses vectors of integers instead of bit vectors as it stores abundances of \kmers.}
    \label{fig:modules}
\end{figure}

Exact structures often demonstrate software modularity. Several examples are depicted in Figure \ref{fig:modules}. As a consequence, some developments have focus on a more restraint aspect, for instance GGCAT~[\cite{cracco2023extremely}] builds de Bruijn graphs, colored de Bruijn graphs and spectrum preserving string sets from \kmers, with no real emphasis on query or compression.

Generalist indexes use these blocks. %For instance, Fulgor [\cite{fan2024fulgor}] and Mac-dBG [\cite{Pibiri2024}] are hash tables that process \kmers into SPSS using the de Bruijn graph builder GGCAT~[\cite{cracco2023extremely}], then creates a MPHF using SSHash~[\cite{pibiri2022sparse}] and compresses color vector. %Mantis [\cite{}] adapts the counting quotient filter Squeakr to record color equivalence classes of \kmers.
For instance, the tool Metagraph can switch between three different data structures according to the needs. One is an indicator bitmap, another is a dynamic hash table and the last is based on a \kmer BWT structure. It also has different color compression strategies. Compression strategies vary across implementations, highlighting a need for standardized binary presence/absence vector formats.

%Early examples include Mantis~[\cite{Pandey2018}], that uses Squeakr (a counting quotient filter that can be exact) to associate \kmers to color equivalence classes, and Reindeer \cite{marchet2020reindeer}, which processes \kmers using BCALM2~[\cite{bcalm2}] for MPHF preparation, stores keys using BLight~[\cite{marchet2021blight}], and compresses abundance vectors with run-length encoding.

\subsection{Compression in exact methods}\label{section:compression}

Most methods include their own strategy for compression. A few articles have been dedicated to compression only, such as \cite{kitaya2021compression} that focus on compressing the graph's sequences with SPSS, \cite{rahman2024compression} that proposes to compress the graph sequences and the colors, and \cite{Pibiri2024} and \cite{mustafa2018dynamic} that focus on color compression.

\subsubsection{Graph compression}
Compression of graph sequences is usually achieved either relying on SPSS [\cite{fan2024fulgor,rahman2024compression}], or on BWT-based compression [\cite{pandey2018mantis, Karasikov2020.10.01.322164, Alanko2023}]. For both approaches, variants make compression worse. BWT-based runs can become shorter, and variation patterns in de Bruijn graph often split common SPSS sequences into shorter strings. Typically, in a bubble structure, one path forms a longer superstring while the other is shorter.

\cite{rahman2024compression} relies on heuristic SPSS to get long superstrings and nests the alternative shorter path associated to a variant within the larger superstring. Additionally, this leverages common color patterns in the variants for more effective color encoding.

\subsubsection{Color compression of binary vectors and integer vectors}
A first solution is to vary compression strategies with vector sparsity. Sparse vectors store ranks of 1s, dense vectors use complements to sparse vectors.  Mid-range vectors use run encoding strategies (RRR, RLE or Roaring bitmaps) that replace consecutive repeated characters  with a single character and a count of its repetitions (see Figure ~\ref{fig:compression_strat} (c), e.g. [\cite{yu2018seqothello, Alanko2023,Pibiri2024}]). Some methods also concatenate all color vectors before applying this type of encoding on a single flat row (e.g. [\cite{pandey2018mantis, Karasikov2020.10.01.322164}]). More advanced run encoding strategies working on both rows and columns of a presence/absence matrix were proposed in the context of colored \kmer sets  ([\cite{karasikov2020sparse}] in \cite{Karasikov2020.10.01.322164}).

$K$-mers in a single column of a \kmer $\times$ color matrix are expected to be extremely sparse, so these schemes have also been used to perform a column-first compression of the colored matrix, for a more severe compression [\cite{Karasikov2020.10.01.322164}]. The counterpart is a slower access during queries.

\subsubsection{Consecutive \emph{k}-mer's colors are correlated}
Color compression leverages similarities between consecutive \kmers, storing differences between vectors. Some methods simply de-duplicate similar color vectors and have every entry point to it, as [\cite{marchet2020reindeer}]. It is also possible to construct SPSS whose \kmers are all associated to the same color vector, and where a representative is stored for all \kmers (e.g. \cite{Alanko2023, marchet2020reindeer}).  Similarly, other algorithms propose to cluster SPSS with similar color to improve color compression [\cite{rahman2024compression}] (see Figure ~\ref{fig:compression_strat} (a)).

Since consecutive \kmers are likely to share presence/absence patterns, it is also possible to encode only the difference between row representing consecutive \kmers, with trade-offs at query time~[\cite{10.1093/bioinformatics/btab330}].

\subsubsection{Color classes have redundant content}

\begin{figure}[ht]
    \centering
    \includegraphics[width=1.1\textwidth]{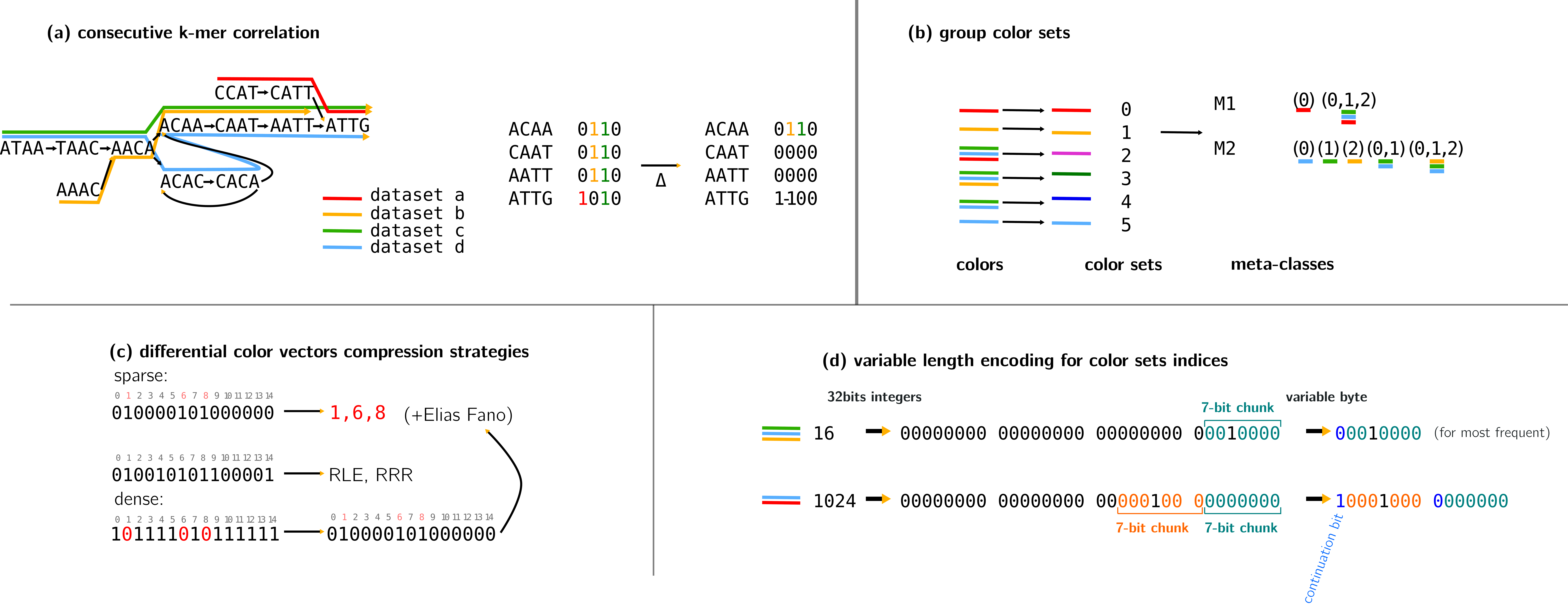}
    \caption{Intuitions of some row first color compression strategies - In (a), keeping a single representative \kmer for mono-color unitigs/SPSS is not pictured. (a) shows how consecutive \kmer can benefit from encoding differences between presence because they are likely to have similar patterns. It is the basic observation behind efficient compressing techniques such as \cite{10.1093/bioinformatics/btab330}. (b) shows how \cite{Pibiri2024} groups color sets. For (c) a more advanced row + columns compression was proposed in \cite{karasikov2020sparse}. Column-first compression does not appear. (d) shows how to move from a fixed length encoding to variable length by dividing the inital representation in 7 bits chunks until no 1 is found. Column-first compression of \cite{Karasikov2020.10.01.322164} is not pictured.}
    \label{fig:compression_strat} 
\end{figure}

The number of color sets can grow exponentially with the number of colors, with recent practical datasets showing over hundreds of thousands of color sets. 
A complementary strategy is therefore to use fewer bits for the integer representation of most frequent color sets (\cite{rahman2024compression, pandey2018mantis}). Often present in implementations but not described in papers, variable length encoding adapts representation of integers, allowing to use variable numbers of bits instead of fixed-sized integer representations (see Figure ~\ref{fig:compression_strat} (d). Other solutions adopt less simple schemes with better compression ratios, such as Huffman encoding [\cite{rahman2024compression}].\\
It is also possible to see vectors of color sets as sparse sequences of positive increasing integers, that can be further compressed using schemes such as Elias-Fano (e.g. in \cite{rahman2024compression}).\\ %todo chercher autres elias fano
Another contribution looked for the minimum number of color classes to represent a colored de Bruijn graph [\cite{alipanahi2018recoloring}].

For quite some time, it has been remarked that \kmers close in the de Bruijn graph had strong correlations in color classes~[\cite{almodaresi2021incrementally}]. Therefore, some color sets are very redundant in their color color content. Recently, meta-classes [\cite{Pibiri2024}] were proposed to encode shared color patterns efficiently when the input is a set of close samples or genomes. This strategy is particularly interesting because it can be paired with the super-\kmer partition used in hash tables ~[\cite{Pibiri2024}]. This type of colored compressed graph have been called Mac-DBG and is currently implemented in Fulgor (see Figure ~\ref{fig:compression_strat} (b)).

\section{Practical considerations}
\subsection{Time, disk, and RAM footprints}

The performance of colored \kmer set structures varies significantly in terms of disk space, memory usage, computation time (both for indexing and querying), and the occurrence of false positives and negatives. Currently, a comprehensive benchmark encompassing all methods is not yet available, but efforts are underway in the Panbench initiative\footnote{\url{https://www.panbench.eu/}}.

\subsubsection{Resources}

Some tools, such as kmIndex, are designed to be disk-intensive, prioritizing scalability. Pantools, on the other hand, requires significant disk space to facilitate extensive interactions with the colored graph through a database. %Scalability can reach up to XXXX in some cases.

Index construction can be particularly time-consuming, especially when complex compression techniques are applied or when multiple successive processing steps are required. This process may also demand more temporary disk space than the final index size itself. Some indexes worked on reducing the RAM footprint by loading only necessary fractions of the index at a time during constrution (e.g. \cite{10.1093/bioinformatics/btad225}) and query (e.g. \cite{seiler2021raptor}).
Conversely, GGCAT focuses on graph construction and is designed to construct unitigs rapidly, emphasizing speed during the indexing phase.

While most indexes need to be rebuilt from scratch when adding or removing datasets, a few tools allow for the dynamic insertion of new datasets (see subsection \ref{section:dynamicity}).

\subsubsection{Query speed}
Query speed is influenced by the cache efficiency and the level of compression applied to \kmers and their associated colors. As the index grows, especially with an increasing number of colors, query speed tends to decrease. The fastest query times are in the range of hundreds of nanoseconds, relying on efficient caching of queried \kmers to accelerate processing. However, more compressed indexes, such as \cite{Karasikov2020.10.01.322164}, may experience slower query speeds due to the additional overhead of decompression.

\subsection{Functionalities}
Some tools are more user-friendly when offered as a service via web servers, such as Reindeer, BIGSI/COBS, Metagraph, Pebblescout and kmIndex. Many tools are also capable of building indexes directly on a laptop, and some, like Bifrost [\cite{holley2019bifrost}], have a large community of users and are supported by external projects.

\subsubsection{I/Os}
Many static data structures, specialized in efficient \kmer queries, require a \kmer set as input (e.g., those relying on minimal perfect hashing), necessitating pre-treatment of raw data. Static structures require the raw data to be converted into a de Bruijn graph (a list of \kmers or unitigs). Therefore, depending on filters applied to raw reads when building input unitigs for instance, exact structures may not contain the exact same \kmer set that is present initially. Conversely, tools specialized in building de Bruijn graphs accept raw data as input and are capable of applying filters and corrections to the dataset. There are less tools in that category (Bifrost, GGCAT and Metagraph).\\
Similarly, Reindeer needs precomputed counts (in the BCALM2 [\cite{bcalm2}] format), while Needle and Metagraph can count \kmer by themselves.

A common limitation is the size of k, often constrained to 31 due to the use of 64-bit integers to represent \kmers. Additionally, most tools treat forward and reverse \kmers as the same unit  and do not support stranded queries. GGCAT and Themisto [\cite{alanko2023themisto}] are an exception, GGCAT allowing to index forward and reverse \kmers separately, Themisto takes the forward/reverse of the \kmer as parsed in the read.

\subsubsection{Membership queries, pseudoalignment, and read alignment}

Colored structures enable mapping query sequences to annotated \kmers without performing full sequence alignment, identifying regions in the reference where the query sequences likely originate based on shared \kmers. For colored \kmer sets, these queries also return the "colors" of the query \kmers, indicating which datasets contain the \kmer. Depending on the tool, it may return colors for each \kmer or the intersection of colors across \kmers in a query (a process known as pseudoalignment, as in ~[\cite{fan2024fulgor,Pibiri2024}]).

Pseudoalignment and \kmer queries are highly efficient, significantly reducing computational overhead compared to traditional alignment methods. However, increasing the number of colors typically leads to higher memory usage and potentially slower query times.

It is also possible to align sequences to de Bruijn graphs using alignment heuristics derived from pairwise sequence alignment [\cite{mustafa2024label}].
One contribution of colored de Bruijn graphs has been the use of colors as additional information to compute color-coherent alignments [\cite{10.1093/bioinformatics/btae226}].

\begin{figure}[ht]
\centering
\includegraphics[width=\textwidth]{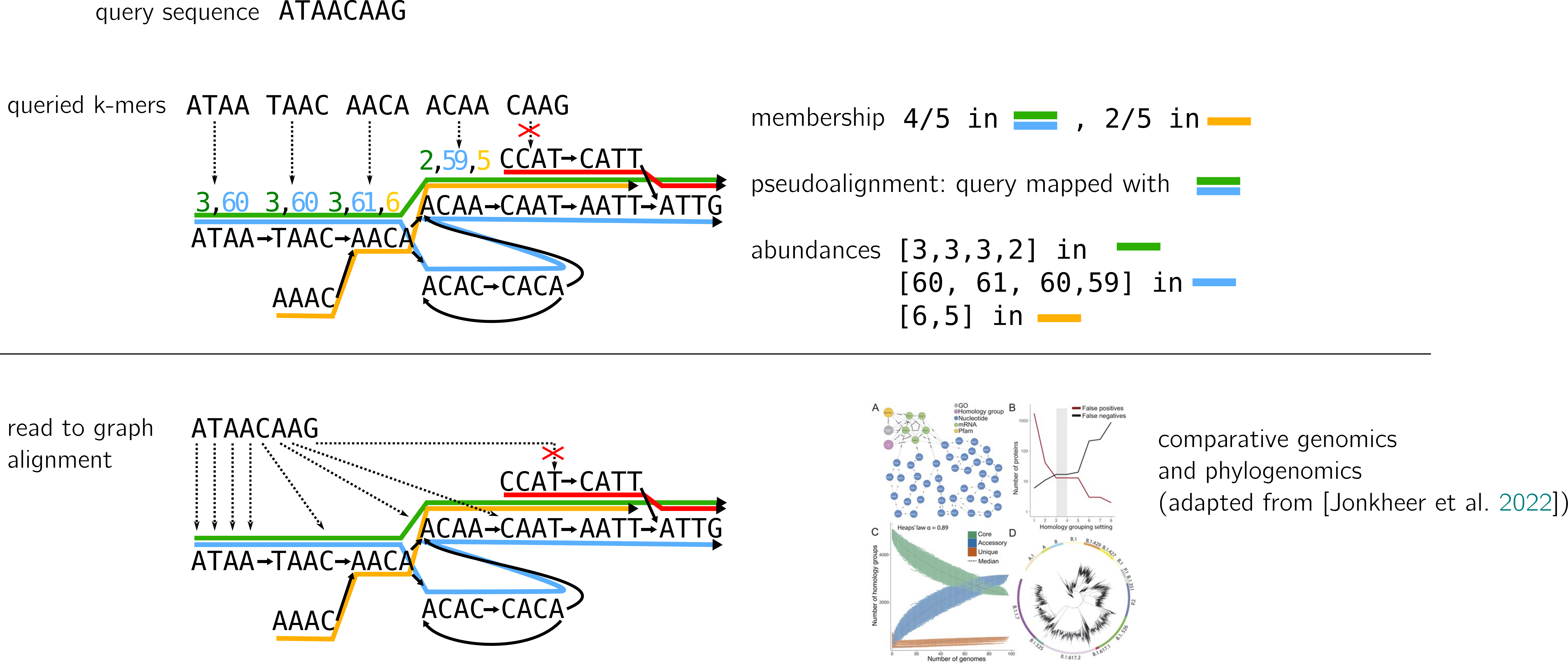}
\caption{Possible queries with colored \kmer sets. Most common queries are on top. For abundance queries, when parsing raw outputs, \cite{bessiere2024exploring} and \cite{darvish2022needle} showed that median of the abundance in a color works best to summarize abundance queries. Read to graph alignement is more time consuming. Pantools offers a full range of advanced manipulation on data from the graph. Other tools embedded in webservices also offer more metadata and visualization possibilities.}
\label{fig:dynamicity_colors}
\end{figure}

\subsubsection{Extended query information}

Beyond simple membership, several methods now allow for querying \kmer abundances, either approximately (Needle, Metagraph) or exactly (Reindeer), across a collection of datasets. Pantools offers a more comprehensive set of tools, such as functional annotation and homology grouping of \kmers. Grimr~[\cite{ingels2024constrained}], for example, enables the extraction of a subset of \kmers from a colored set, meeting specific constraints (e.g., appearing in samples 1 and 2 of the colored set but not in samples 3 and 4).

%TOOD query size

\subsection{Dynamicity}\label{section:dynamicity}
Since most exact methods integrate static \kmer data structures, dynamicity (the ability to insert \kmers) has traditionally been achieved through buffering strategies that rebuild the entire index from smaller ones. For instance, VARI-Merge~[\cite{muggli2019building}] merges several graph BWT structures for its colored \kmer set. Bifrost [\cite{holley2019bifrost}] allows direct updates to the graph's unitigs. Recent advancements have introduced more dynamic capabilities. For example, the updated version of Mantis [\cite{Almodaresi2022}] supports incremental updates by merging new data into existing indexes without full reconstruction. Similarly, dynamic compression can also be applied to colors, as explored in [\cite{10.1093/bioinformatics/bty632}].

Inexact structures seem more adequate for insertions since at first glance they require adding a leaf to a tree or a column to the matrix, however in practice this involves developments (update of tree's inner nodes, inverted matrix) not yet present in the implementations, except for PAC.

\section{Applications}
Applications of colored \kmer sets include large-scale query web services and specialized indexes.

\subsection{All clades}
%number of color classes ? blackwell 3000000

Using the Metagraph structure, \cite{Karasikov2020.10.01.322164} developed an indexing framework for efficiently managing and querying vast genomic datasets. The system integrates taxonomic and functional annotations, supporting diverse genomics research applications. By providing a scalable infrastructure, the framework enables comprehensive analysis across a wide array of biological sequences\footnote{\url{https://metagraph.ethz.ch/}}, with 1.9 million datasets and over 4.8 Petabases ($\approx$ 2.5 Petabytes) as of today.
%TODO ordre de gdeur

An instance of Pebblescout is hosted by the NCBI\footnote{\url{https://pebblescout.ncbi.nlm.nih.gov/}} and allows queries in over 3.7 Petabases of SRA data, in RefSeq assembled genomes, SRA metagenomics, metatranscriptomics and human RNA-seq runs, as well as microbial samples. It outputs a list of projects and runs where the output has matches.

\subsection{Metagenomes}
 KMCP [\cite{10.1093/bioinformatics/btac845}] is a \kmer based metagenomic profiling tool designed for accurate taxonomic profiling of both prokaryotic and viral populations from metagenomic shotgun sequence data, using a matrix-like index. Not technically a colored de Bruijn graph but closely related, Kraken2 [\cite{wood2019improved}] uses a sampling approach based on minimizers associated to a hash table to efficiently pair sequences with their lowest common ancestor, and achieves fast metagenomic classification.
 
 In \cite{alipanahi2020metagenome}, in order to track organisms and genes linked to antibiotic resistance in a microbiome, a colored de Bruijn graph is used as a support for SNP calling in antibiotic resistance genes from metagenomics data. In that case, "colors" are read provenance of a \kmer.

Based on memory-mapped Bloom filters, the Ocean Read Atlas (ORA) [\cite{Lemane2024}], is a web server enabling real-time queries on over a thousand of ocean metagenomics datasets from the Tara Oceans project\footnote{\url{https://tara-oceans.mio.osupytheas.fr/}}.
It enables efficient exploration of oceanic microbial diversity and gene functions. In human metagenomics, \cite{almeida2021unified} have generated a BIGSI of their Unified Human Gastrointestinal Genome collection, comprising over 204,000 non redundant genomes from more than 4,000 gut prokaryotes. The index allows users to interactively query the collection.

\subsection{Phylogenomics, pangenomics}
The bacterial genome analysis field motivated original advancements using colored de Bruijn graphs. For instance, ggCaller [\cite{horsfield2023accurate}] is tool for pangenome annotation and clustering using a colored de Bruijn graph represented with Bifrost. Combining gene prediction, functional annotation, and clustering using population-wide de Bruijn graphs, ggCaller enhances gene prediction accuracy and orthologue clustering.

PAN-GWES [\cite{kuronen2023pangwes}] is a method for discovering co-selected and epistatically interacting genomic variations, also based on colored de Bruijn graphs. This approach identifies associations between loci linked with drug resistance and adaptation in pathogens, providing a phenotype- and alignment-free method for analyzing bacterial adaptation and evolution in large populations.

Using a colored de Bruijn graph from genomic data, the SANS approach [\cite{wittler2020alignment}] extracts common subsequences to infer phylogenetic splits. The approach avoids pairwise comparisons and alignment, using unitigs to identify separation signals among genomes. Pantools~[\cite{10.1093/bioinformatics/btw455}] leverages the colored de Bruijn graph to allow functional annotation, analysis and exploration of pangenomes.

The literature also offers queryable large indexes. A few years ago, \cite{Luhmann2021} used Bifrost to conceive a tool for efficiently querying hundreds of thousands of bacterial genome assemblies. \cite{Brinda2023.04.15.536996} introduces phylogenetic compression, that works by using evolutionary history to group and reorder microbial genomes in a way that maximizes local similarities between related genomes. This technique improves data compression ratios for genome assemblies, de Bruijn graphs, and \kmer indexes. Notably the tool allows membership queries on compressed COBS indexes, for several large microbial banks\footnote{\url{brinda.eu/mof}}. This year, with the help of this technique, Martin Hunt and colleagues introduced "AllTheBacteria" [\cite{Hunt2024.03.08.584059}], a platform compiling all known bacterial genomes\footnote{\url{https://github.com/AllTheBacteria/AllTheBacteria}}, a step further after their release of 661 thousands assembled bacterial genomes [\cite{10.1371/journal.pbio.3001421}]. The platform integrates advanced assembly techniques and efficient search algorithms, facilitating quick retrieval of genomic information. 

\subsection{RNA-seq}
The Transipedia initiative [\cite{bessiere2024exploring}] explores RNA isoform diversity in a large cancer cell line of over one thousand RNA-seq datasets. Uusing the colored index Reindeer, it allows to associate \kmers to their abundances. This study demonstrates the application of \kmer sets for querying RNA sequences, providing insights into the transcriptomic landscape of cancer cell lines and aiding in the identification of novel RNA isoforms and potential biomarkers. It also provides a platform to explore different cancer datasets and their metadata\footnote{\url{https://transipedia.fr/}}.\\
Thanks to the reference-free property of colored de Bruijn graphs, Reindeer helped identifying a novel splicing variant for the tumor suppressor PDCD4, that translates into a smaller protein, putativley less active in its suppressor function [\cite{corre2024}].

\section{Summary, trends and directions for colored \emph{k}-mer sets}

In this review we saw both tools specialized on constructing colored de Bruijn graphs and tools specialized in building indexes for membership queries based on colored de Bruijn graphs. There exist a continuum between methods based on sampling such as those based on MinHash, inexact methods and exact methods to index colored \kmer sets, trading precision for space in methods allowing errors. Large queries are therefore the normal use case for these structures, whereas exact methods are necessary for applications demanding precise results, sometimes at the \kmer resolution.

\paragraph{\textbf{Better benchmarks and scalability}} A comprehensive study on the practical scalability and resource usage with respect to the number of samples is essential but not yet achieved. Similarly, the complexity of datasets can significantly affect resource usage, particularly for \kmer indexes based on lexicographic properties, which may experience performance degradation in the presence of increased errors or mutations. Current methods have reached several Petabytes of data through highly compressed schemes [\cite{Karasikov2020.10.01.322164}], and reached 100,000 RNA-seq samples for quick exploration [\cite{10.1093/bioinformatics/btad225}]. It is to be expected that these indexes will continue growing, especially given the renewed accessibility of SRA data through project Logan [\cite{chikhi2024logan}], that gives access to SRA data as assembled units.

\paragraph{\textbf{Novel operations}}  Some indexes can now support insertions and can provide additional information beyond simple presence/absence data. While they are marginal, advancement in compression of this new types of information should make them more common in the future. These advancements will also promote indexes associating \kmers to various types of information or metadata (new types of query include read of origin, and working with long reads as in \cite{vandamme2024tinted}). 

%data availability for a good benchmark and biol validation

%These changes will be easier to implement, from engineering-management and economic points of view, if they occur within big system components: reusable software with typically more than a million lines of code or hardware of comparable complexity. When a single organization or company controls a big component, modularity can be more easily reengineered to obtain performance gains. 

\paragraph{\textbf{Software development}} The structures described hereby are strongly tied to software development. Some are proofs of concepts, other aim at being widely used. 
Methods have been more and more carved while taking hardware into account. Methods preserving cache-locality have such as fast query that they  allow to leave only the server's processing time in recent web-indexes. Memory-mapped indexes proved not to sacrifice too much query speed and reduce memory footprints during construction and querying, leveraging the performance of SSDs.\\
Many new methods include task specialization through modular implementations, a trend that could accentuate, helped with the rise of Rust and its helpful package manager. Modularity has several advantages: performance gains occur from combination of different modules, and it makes easier for other groups to access and re-engineering parts of a tool.

\paragraph{\textbf{Dissemination \& outreach}} All these structures remain quite addressed to specialists. They work in command line and involve some knowledge on the parametrization and on the underlying structures to interpret the outputs. To make these tools more accessible to a broader community, several steps could be taken. Firstly, recent papers that showcase their usage, parameters, and capabilities across various datasets can serve as practical guides.
Indeed, these methods require careful query engineering and can greatly benefit from user expertise. For example, \cite{bessiere2024exploring} demonstrated that when searching for mutated \kmers, it is most effective to include only those \kmers that specifically cover the mutation in the query. More broadly, there is considerable potential for improvement in querying \kmers that are expected to differ from the indexed content due to mutations.\\
Additionally, the development of intuitive interfaces for querying and interpreting outputs, would make these tools more approachable for users. With their patterns linked to genomic and RNA variations, de Bruijn graphs offer valuable and scalable data visualizations, but this aspect is yet underdeveloped.

\paragraph{\textbf{Other sequences}} A related future direction will be to index and query amino acids or different alphabets. For proteins, this functionality starts to appear from time to time in the literature [\cite{10.1093/bioinformatics/btad101, luebbert2023efficient, Karasikov2020.10.01.322164}], but vastly remains to be explored. Other alphabet could include higher order de Bruijn graphs for long read assembly, for instance relying on an alphabet of minimizers instead of bases, as explored in [\cite{bankevich2022multiplex,benoit2024high}].

\section*{Fundings}

This study has been supported by ANR JCJC Find-RNA [ANR -23-CE45-0003-01].

\section*{Conflict of interest disclosure}

The author declares that she complies with the PCI rule of having no financial conflicts of interest in relation to the content of the article.

\section*{Data, script, code, and supplementary information availability}
None declared.

\section*{Acknowledgments}
This manuscript and its companion were created using my notes from talks I gave in recent conferences, courses and workshops. I'd like to thank the community for inviting me and giving me a opportunity to present and discuss my views on those subjects. I'd also like to thank J. Alanko, G. Pibiri and L. Robidou for going over the manuscript.

%\bibliographystyle{alpha}
%\bibliography{PCJ-sample}
\bibliography{arxiv}

\end{document}